\documentclass[prc,onecolumn,showpacs,superscriptaddress]{revtex4}
\usepackage{graphicx}

\begin{document}

\title{Predictions of the FBD model for the synthesis
cross sections of $Z$ = 114-120 elements based on
macroscopic-microscopic fission barriers }

\author{K. Siwek-Wilczy\'nska}
\affiliation{Institute of Experimental Physics, Faculty of
Physics, University of Warsaw, Ho\.za 69, 00-681 Warsaw, Poland }

\author{ T. Cap}
\affiliation{Institute of Experimental Physics, Faculty of
Physics, University of Warsaw, Ho\.za 69, 00-681 Warsaw, Poland }

\author{M. Kowal}
\affiliation{National Centre for Nuclear Research, Ho\.za 69,
00-681 Warsaw, Poland   }

\author{A. Sobiczewski}
\affiliation{National Centre for Nuclear Research, Ho\.za 69,
00-681 Warsaw, Poland   }

\author{J. Wilczy\'nski}
\affiliation{National Centre for Nuclear Research, 05-400
Otwock-\'Swierk, Poland   }


\begin{abstract}

A complete set of existing data on hot fusion reactions leading to
synthesis of superheavy nuclei of $Z$ =114-118, obtained in a
series of experiments in Dubna and later in GSI Darmstadt and LBNL
Berkeley, was analyzed in terms of a new angular-momentum
dependent version of the fusion-by-diffusion (FBD) model with
fission barriers and ground-state masses taken from the Warsaw
macroscopic-microscopic model (involving non-axial shapes) of
Kowal et al. The only empirically adjustable parameter of the
model, the injection-point distance ($s_{inj}$), has been
determined individually for all the reactions. Very regular
systematics of this parameter have been established. The
regularity of the obtained $s_{inj}$ systematics indirectly points
at the internal consistency of the whole set of fission barriers
used in the calculations. (In an attempt to fit the same set of
data by using the alternative theoretical fission barriers of
M\"oller {\em et al.} we did not obtain such a consistent result.)
Having fitted all the experimental excitation functions for
elements $Z$ = 114--118, the FBD model was used to predict cross
sections for synthesis of elements $Z$ = 119 and 120. Regarding
prospects to produce the new element $Z$ = 119, our calculations
prefer the $^{252}$Es($^{48}$Ca,xn)$^{300-x}$119 reaction, for
which the synthesis cross section of about 0.2 pb in $4n$ channel
at $E_{c.m.}\approx 220$ MeV is expected. The most favorable
reaction to synthesize the element $Z$ = 120 turns out to be
$^{249}$Cf($^{50}$Ti,xn)$^{299-x}$120, but the predicted cross
section for this reaction is only 6 fb (for $3n$ and $4n$
channels).
\end{abstract}



\maketitle

\section{Introduction}

Superheavy nuclei of $Z \ge 104$ were synthesized either in
``cold'' fusion reactions on closed-shell $^{208}$Pb and
$^{209}$Bi target nuclei bombarded by projectiles ranging from Ti
to Zn or in "hot" fusion reactions, in which the heaviest
available actinide targets were bombarded with the neutron rich
$^{48}$Ca projectiles. See review articles
\cite{Hofmann-review,Hof-Munz-review} and \cite{Oganessian},
respectively. In the cold fusion reactions only one neutron is
emitted from the compound nucleus to form the final
compound-residue nucleus in its ground state. In hot fusion
reactions more neutrons are emitted. At each step of the
deexcitation cascade the neutron evaporation competes with the
dominating process of fission. Therefore the synthesis cross
section represents only a small part of the fusion cross section.

A characteristic feature of the fusion-evaporation reactions
leading to the synthesis of superheavy nuclei is enormous
hindrance of the fusion process itself. Consequently, the cross
sections for the synthesis of heaviest elements are measured in
picobarns or even femtobarns. It is believed that the hindrance is
caused by the highly dissipative dynamics of the fusing system in
its passage over the saddle point on the way through the
multidimensional potential energy surface from the initial
configuration of two touching nuclei into the configuration of the
compound nucleus. Zagrebaev and Greiner developed a method of
solving Langevin equations of motion to describe this stochastic
stage of the fusion process \cite{Zagreb-Langevin}. In spite of
very time consuming Langevin trajectory calculations, in which
only one of say a million trajectories leads to formation of the
compound nucleus, the model is used effectively to calculate
synthesis cross sections for various reactions \cite{Zagreb-08}.
Another approach to the process of fusion of a ``dinuclear
system'' (DNS) was proposed in Ref. \cite{DNS}. It was assumed in
this model that the dinuclear system stays in contact
configuration and undergoes successive transfer of all nucleons
from the lighter nucleus to the heavier partner (in competition
with the quasi-fission processes). Applications of this concept
have been used in recent years by several groups. In still another
approach, the ``fusion-by-diffusion'' model \cite{FBD-05}, the
stochastic process of shape fluctuations that lead to the
overcoming the saddle point was described as the solution of the
Smoluchowski diffusion equation in the deformation space along the
fission valley.

The cold fusion reactions leading to the synthesis of nuclei of $Z
\le 113$ were studied systematically with the DNS model in Ref.
\cite{Cherepanov}, with the fusion-by-diffusion (FBD) model
\cite{FBD-05}, \cite{FBD-11} and with the Langevin dynamics model
\cite{Zagreb-08}. The hot fusion reactions leading to the
synthesis of the heaviest nuclei of $Z \ge 114$ have not been
studied so systematically. In Ref. \cite{Zagreb-08} excitation
functions for some selected reactions were calculated although
they were not confronted with experimental cross sections. Most of
the publications on this topic concentrated on the predictions
concerning possible ways of synthesis of the heaviest elements of
$Z=119$ and 120 \cite{Zagreb-08,Liu-120,Adamian-2009,Siwek-120,
Nasirov-120,Nasirov-120a,Zagreb-12,Wang-12}. Only very recently,
an extensive study of cold and hot fusion reactions in terms of a
phenomenological approach based on the DNS model was reported
\cite{Scheid-syst}.

There is one important aspect of all the models of the synthesis
of superheavy nuclei that was not treated with proper attention so
far. This is the question of the choice of theoretical fission
barriers and ground-state masses which have to be adopted for the
description of the deexcitation of the compound nucleus. It is
well known that calculations of the cross sections for synthesis
of superheavy nuclei are extremely sensitive to the height of the
fission barrier, especially in case of ``hot'' fusion reactions,
in which three or four neutrons are emitted from the compound
nucleus. When the barrier heights are not known precisely, an
error in evaluation of the $\Gamma_n/\Gamma_f$ ratio in each step
of the ($xn$) deexcitation cascade accumulates $x$ times leading
to enormous errors in the calculation of the synthesis cross
sections. (Here, $\Gamma_n$ and $\Gamma_f$ denote the neutron
decay width and fission width, respectively.) Thus, precise
knowledge of theoretical fission barriers and neutron binding
energies (ground-state masses) is crucial for reasonable
predictions of the synthesis cross sections.

In the last decade the mass tables of M\"oller et al.
\cite{Moller} have most frequently been used in the field of
superheavy nuclei. Unfortunately, fission barriers heights are not
given in these tables. Therefore, in most of the mentioned above
calculations of the synthesis cross sections the ground-state
shell effect of the compound nucleus (that is listed in these
tables) was used as the barrier height. In this simplification,
both the macroscopic deformation energy and the shell effect at
the saddle configuration are neglected. It seems, therefore, that
these approximate values of the fission barrier are not
sufficiently accurate to guarantee reliable predictions of the
synthesis cross sections. (Absolute value of both these neglected
effects may be of about 1--2 MeV each, while a 1 MeV-shift of the
barrier height may result in a change of the calculated cross
section of $3n$ or $4n$ reaction by 2--3 orders of magnitude.)

Only in recent years systematic compilations of theoretical
fission barriers of superheavy nuclei (combined with the necessary
information on the ground-state masses) have became available in
literature. Calculations in framework of the macro-microscopic
approach were reported by Muntian et al. \cite{Muntian} and later
by M\"oller at al. \cite{Moller-barriers}. The model
\cite{Muntian} has been extended recently by Kowal et al.
\cite{Kowal-10,Kow-Sob} by the inclusion of nonaxiality as an
important new degree of freedom. Fission barriers of superheavy
nuclei have been calculated also in a number of other papers
within various models (see Ref. \cite{Ring}, Table IV for a
review), however no sufficiently systematic information on the
fission barriers and, simultaneously, ground-state masses has been
provided.

In the present study we adopt the ``fusion-by-diffusion'' (FBD)
model \cite{FBD-05,FBD-11} for calculating the synthesis cross
sections of the heaviest nuclei in hot fusion ($xn$) reactions by
using the information on the fission-barrier heights
\cite{Kowal-10,Kow-Sob} and other properties of the superheavy
nuclei obtained within the Warsaw macroscopic-microscopic model
\cite{Muntian}.

The whole set of experimental data
\cite{Oganessian,Oga-243Am,Oga-244Pu,Oga-242Pu,Oga-249Cf,LBL-242Pu1,Oga-249Bk,GSI-244Pu,LBL-242Pu2,GSI-248Cm,GSI-244Pu2,Oga-243Am2}
on the synthesis of new superheavy elements of $Z$ = 114--118
(obtained in Dubna by Oganessian and coworkers and later in a
series of confirming experiments at GSI Darmstadt and LBNL
Berkeley) was analyzed. Based on this test of the model
predictions, the calculations were then performed for
experimentally unexplored yet reactions aimed at the synthesis of
new elements of $Z$ = 119 and 120.

\section{Review of the FBD model}

The fusion-by-diffusion (FBD) model \cite{FBD-05,FBD-11} serves
for calculating cross sections for the synthesis of superheavy
nuclei. Recently the model was modified in order to describe both
cold fusion ($1n$) and hot fusion ($xn$) reactions. In this
extended version \cite{FBD-11}, for each angular momentum $l$ the
partial evaporation-residue cross section $\sigma_{ER}(l)$ for
production of a given final nucleus in its ground state is
factorized as the product of the partial capture cross section
$\sigma_{cap}(l)=\pi \lambdabar^2(2l+1)T(l)$, the fusion
probability $P_{fus}(l)$ and the survival probability
$P_{surv}(l)$:
\begin{equation}
\label{factorize}
 \sigma_{ER} = \pi \lambdabar^2 \sum_{l = 0}^{\infty}(2l+1)
 T(l)\cdot P_{fus}(l)\cdot P_{surv}(l).
\end{equation}
The capture transmission coefficients $T(l)$ are calculated in a
simple sharp cut-off approximation, where the upper limit
$l_{max}$ of full transmission, $T(l)=1$, is determined by the
capture cross section known from the systematics described in
Refs. \cite{FBD-11,KSW04}. Here $\lambdabar$ is the wave length,
$\lambdabar^2=\hbar^2/2\mu E_{c.m.}$, and $\mu$ is the reduced
mass of the colliding system. The fusion probability $P_{fus}(l)$
is the probability that the colliding system, after reaching the
capture configuration (sticking), will eventually overcome the
saddle point and fuse, thus avoiding reseparation. The other
factor in Eq. (1), the survival probability $P_{surv}(l)$, is the
probability for the compound nucleus to decay to the ground state
of the final residual nucleus via evaporation of light particles
and $\gamma$ rays, thus avoiding fission.

The cross sections for the synthesis of superheavy nuclei are
dramatically small because the fusion probability $P_{fus}(l)$ is
hindered (in some reactions even by several orders of magnitude)
due to the fact that the saddle configuration of the heaviest
compound nuclei is much more compact than the configuration of two
colliding nuclei at sticking. It is assumed in the FBD model that
after the contact of the two nuclei, a neck between them grows
rapidly at an approximately fixed mass asymmetry and constant
length of the system. This ``neck zip'' is expected to carry the
system towards the bottom of the asymmetric fission valley. This
is the ``injection point'', from where the system starts its climb
uphill over the saddle in the process of thermal fluctuations in
the shape degrees of freedom. Theoretical justification of the
above picture of fast zipping the neck was given in Ref.
\cite{Boilley}, where the later stage of the stochastic climb
uphill was described by solving the two-dimensional Langevin
equation. Theoretical location of an effective injection point can
be deduced from this model \cite{Boilley}. Also in a modified
fusion-by-diffusion model \cite{Xe+Xe} the location of the
injection point was estimated theoretically. In our model we rely,
however, on empirical determination of the injection point. Its
location in the asymmetric fission valley, $s_{inj}$, is the only
adjustable parameter of the FBD model.

By solving the Smoluchowski diffusion equation, it was shown in
Ref. \cite{Acta} that the probability of overcoming a parabolic
barrier for the system injected on the outside of the saddle point
at an energy $H$ below the saddle is:
\begin{equation}
\label{hindrance} P_{fus} = \frac{1}{2}(1-{\rm erf}\sqrt{H/T}\,),
\end{equation}
where $T$ is the temperature of the fusing system. The energy
threshold $H$ opposing fusion in the diffusion process is thus the
difference between the energy of the saddle point $E_{saddle}$ and
the energy of the combined system at the injection point
$E_{inj}$, where $E_{inj}$ is calculated using algebraic
expressions given in Ref. \cite{FBD-11} which approximate the
potential energy surface along the fission valley. The energy of
the saddle point is given by the adopted theoretical value of the
fission barrier $B_f$ and the ground-state energy of the compound
nucleus. The corresponding values of the rotational energy at the
injection point and at the symmetric saddle point are calculated
assuming the rigid-body moments of inertia at these configurations
\cite{FBD-11}.

As regards the survival probability $P_{surv}$, the standard
statistical-model calculation were done by applying the Weisskopf
formula for the particle (neutron) emission width $\Gamma_n$, and
the conventional expression of the transition-state theory for the
fission width $\Gamma_f$. The level density parameters $a_n$ and
$a_f$ for neutron evaporation and fission channels were calculated
as proposed by Reisdorf \cite{Reisdorf}, with shell effects
accounted for by the Ignatyuk formula \cite{Ignatyuk}. All details
regarding the calculations of the survival probability $P_{surv}$
can be found in our recent paper \cite{FBD-11}. In case of
calculating multiple evaporation ($xn$) channels a simplified
algorithm avoiding the necessity of using the Monte Carlo method
was applied \cite{Cap-xn}.

\section{Calculations for Z=114-120 elements with the
macro\-scopic-microscopic barriers}

As pointed out in the Introduction, calculations of the cross
sections for synthesis of superheavy nuclei are extremely
sensitive to the height of the fission barrier, especially in case
of ``hot fusion'' reactions because at each step of deexcitation
cascade the competition between neutron emission and fission
strongly depends on the difference of energy thresholds for these
two decay modes. Therefore, in attempts to reasonably calculate
the synthesis cross sections, the choice of realistic and
consistent theoretical information on the fission barrier heights
and the ground-state masses is essential. In our previous
applications of the FBD model, devoted mostly to analysis of cold
fusion reactions (of $Z$ of the compound nucleus $Z_{CN}\le 113$),
fission barriers based on the Thomas-Fermi model
\cite{Thomas-Fermi} were used. In Ref. \cite{KSW-09} it was
observed, however, that for heavier nuclei of $Z_{CN} \ge 114$
produced in hot fusion reactions the fission barriers based on the
Thomas-Fermi model are evidently too high, while barriers based on
the Warsaw macroscopic-microscopic model \cite{Muntian} lead to
better agreement with experimental observations. Therefore results
of the new macroscopic-microscopic calculations of the Warsaw
group \cite{Kowal-10}, involving an extended multi-dimensional
deformation space, have been chosen as the saddle-point and
ground-state input to the FBD model. The published \cite{Kowal-10}
results for even-even nuclei  have been supplemented with
unpublished yet results for odd-$Z$ and/or odd-$N$ nuclei
\cite{Kow-Sob}.

In the first stage of calculations a complete set of experimental
data
\cite{Oganessian,Oga-243Am,Oga-244Pu,Oga-242Pu,Oga-249Cf,LBL-242Pu1,Oga-249Bk,GSI-244Pu,LBL-242Pu2,GSI-248Cm,GSI-244Pu2,Oga-243Am2}
 on the synthesis of $Z$=114-118 elements in
reactions induced by $^{48}$Ca projectiles on $^{242,244}$Pu,
$^{243}$Am, $^{245,248}$Cm, $^{249}$Bk and $^{249}$Cf targets was
analyzed with the aim to determine location of the injection point
$s_{inj}$. Here $s_{inj}$ is defined as the excess of the total
length of the combined system over the length of the initial
system (at the touching configuration) when the neck-zip process
brings the system to the asymmetric fission valley.

In order to determine systematics of $s_{inj}$ for the set of hot
fusion reactions
\cite{Oganessian,Oga-243Am,Oga-244Pu,Oga-242Pu,Oga-249Cf,LBL-242Pu1,Oga-249Bk,GSI-244Pu,LBL-242Pu2,GSI-248Cm,GSI-244Pu2,Oga-243Am2},
the individual values of $s_{inj}$ were deduced for each reaction
and each particular xn channel by adjusting the assumed
$s_{inj}$-value to the experimental synthesis cross section at the
maximum of a given xn excitation function. The compilation of so
deduced $s_{inj}$-values is displayed in Fig. 1 as a function of
the kinetic energy excess $E_{c.m.}-B_0$ above the Coulomb barrier
$B_0$. (For the definition of $B_0$ see Ref. \cite{FBD-11}.)

It should be commented here that values of $s_{inj}$ are inferred
from the synthesis cross sections in a model-dependent way,
assuming particular ground-state masses and fission barriers.
Therefore the result of this procedure obviously depends to some
extent on these theoretical input data used in the calculations.
Consequently, the systematics of $s_{inj}$ obtained in
calculations employing different sources of the theoretical input
data may appear different (cf. the $s_{inj}$ systematics obtained
in recent calculations of cold fusion reactions \cite{FBD-11}
analyzed assuming masses and fission barriers based on the
Thomas-Fermi model \cite{Thomas-Fermi}).

It is clearly seen from Fig. 1 that the injection distance
$s_{inj}$ increases with the decreasing energy $E_{c.m.}-B_0$, in
agreement with expectations based on the dynamics of
nucleus-nucleus collisions, for example the classical trajectory
calculations \cite{Feldmeier}. Very good correlation between the
$s_{inj}$-values and the corresponding energies $E_{c.m.}-B_0$ can
be viewed as an argument in favor of the fission barriers of Kowal
et al. \cite{Kowal-10,Kow-Sob} because such a striking correlation
would be very unlikely if the theoretical barrier heights were
inconsistent with experimental values.

A linear fit to the dependence of $s_{inj}$ on $E_{c.m.}-B_0$ in
Fig. 1,
\begin{equation}
s_{inj}\approx 4.09 \ {\rm fm}- 0.192(E_{c.m.}-B_0) \ {\rm
fm/MeV}, \label{sinj}
\end{equation}
represents the only empirical input to our model and once this
systematics of the injection-point distance is determined in form
of Eq. (\ref{sinj}), one can use the FBD model to calculate
excitation functions of fusion-evaporation reactions without any
adjustable parameters.

In Fig. 2 we present a comparison of our FBD model predictions of
excitation functions for different $xn$ channels with experimental
synthesis cross sections (assigned to the corresponding xn
channels) in the following hot fusion reactions:
$^{244}$Pu($^{48}$Ca,xn)$^{292-x}$114
\cite{Oganessian,Oga-244Pu,Oga-242Pu,GSI-244Pu,GSI-244Pu2},
$^{243}$Am($^{48}$Ca,xn)$^{291-x}$115
\cite{Oganessian,Oga-243Am,Oga-243Am2},
$^{245}$Cm($^{48}$Ca,xn)$^{293-x}$116 \cite{Oganessian,Oga-244Pu},
$^{248}$Cm($^{48}$Ca,xn)$^{296-x}$116
\cite{Oganessian,Oga-242Pu,GSI-248Cm},
$^{249}$Bk($^{48}$Ca,xn)$^{297-x}$117 \cite{Oga-249Bk} and
$^{249}$Cf($^{48}$Ca,xn)$^{297-x}$118 \cite{Oganessian,Oga-249Cf}.
The largest deviations of our general fit to the data approach a
factor of 10 that corresponds effectively to a difference of about
0.5 MeV in the assumed height of the theoretical fission barrier.
Given this high sensitivity of the model predictions to the
assumed fission barrier heights, the overall agreement between the
FBD predictions and measured cross sections is quite satisfactory.
(It is rather unlikely that the accuracy of the theoretical
predictions of individual fission barriers might be much better
than $\pm 0.5$ MeV.)

It is instructive to compare results of calculations presented in
Figs. 1 and 2 with predictions for an alternative set of
theoretical fission barriers. In Fig. 3 we present individual
values of the injection distance $s_{inj}$ deduced for the same
set of data on hot fusion reactions
\cite{Oganessian,Oga-243Am,Oga-244Pu,Oga-242Pu,Oga-249Cf,LBL-242Pu1,Oga-249Bk,GSI-244Pu,LBL-242Pu2,GSI-248Cm,GSI-244Pu2,Oga-243Am2},
but obtained assuming fission barriers of M\"oller et al.
\cite{Moller-barriers}, the only alternative, complete set of
necessary information available in literature. The barriers of
M\"oller et al. are considerably higher than barriers of Kowal et
al. \cite{Kowal-10,Kow-Sob}, thus resulting in larger values of
the calculated survival probability $P_{surv}$. Consequently, the
procedure of ``calibrating'' the individual $s_{inj}$ values by
fitting the predictions to experimental cross sections resulted in
larger values of the determined injection distance $s_{inj}$.
Contrary to the consistent systematics of $s_{inj}$ values shown
in Fig. 1, Fig. 3 demonstrates the evident inconsistency of the
set of $s_{inj}$ values obtained for the barriers of Ref.
\cite{Moller-barriers}. It is seen from Fig. 3 that the $s_{inj}$
values range from 5.5 fm to 8.5 fm and are too large to have a
reasonable physical meaning. (In most cases, they correspond to
the injection distance that exceeds the distance of the scission
configuration.) Most importantly, the individual points in Fig. 3
seam to be almost randomly scattered and do not show any
correlation with energy.

There is one more inconsistency that can be noticed when the
fission barriers of Ref. \cite{Moller-barriers} and the
ground-state masses \cite{Moller} are used. Namely, for these high
fission barriers and corresponding $Q$-values, the predicted
positions of the maxima of the $xn$ excitation functions are
shifted by some 5--7 MeV toward lower energies as compared with
the data (and also with respect to the predictions for barriers of
Ref. \cite{Kowal-10,Kow-Sob}). This effect is illustrated in Fig.
4, where the data for $3n$ and $4n$ channels in the
$^{243}$Am($^{48}$Ca,xn)$^{291-x}$115 reaction are compared with
the excitation functions calculated for these two reaction
channels. This considerable energy shift, seen also for other
reactions, stems from the fact that for the M\"oller's barriers
\cite{Moller-barriers} and the corresponding ground-state masses
\cite{Moller}, the fission barrier $B_f$ is larger than the
neutron binding energy $B_n$ for all the compound nuclei formed in
the studied reactions. Consequently, the $\Gamma_n/\Gamma_f$ ratio
rises very fast at low excitation energies thus influencing the
position and shape of the $xn$ excitation functions.

From Figs. 1 and 2 it is seen that contrary to the generally
higher fission barriers of Ref. \cite{Moller-barriers}, the input
data of Kowal et al. \cite{Kowal-10,Kow-Sob} give a reasonable
agreement of the calculated and measured cross sections as well as
the very clear correlation between $s_{inj}$ and $E_{c.m.}-B_0$
that ``calibrates'' the injection distance $s_{inj}$. This
entitles us to believe that the set of theoretical fission-barrier
heights and ground-state masses \cite{Kowal-10,Kow-Sob} is quite
adequate for a wide range of the heaviest nuclei considered in
this study. Therefore we are going to use them  for predictions of
cross sections of yet unexplored reactions aimed at the synthesis
of new elements $Z$ = 119 and 120.

Regarding possibilities to produce the element $Z$ = 119 we
consider, first of all, the most preferred reactions induced by
the favorable beam of $^{48}$Ca on two isotopes of einsteinium,
$^{252}$Es and $^{254}$Es. These extremely difficult-to-produce
targets might possibly be available in the near future. Therefore
we present in Figs. 5(a) and 5(b) the predicted energy dependence
of the $xn$ cross sections in reactions on these two isotopes. The
largest cross section, which turns out to be at the edge of
experimental possibilities (about 0.2 pb in 4n channel at
$E_{c.m.}\approx 220$ MeV), is predicted for the
$^{252}$Es($^{48}$Ca,xn)$^{300-x}$119 reaction. Surprisingly, the
cross section in the reaction on a more neutron-rich target,
$^{254}$Es($^{48}$Ca,xn)$^{302-x}$119, is by one order of
magnitude lower (only about 15 fb). This is a consequence of lower
fission barriers \cite{Kowal-10,Kow-Sob} in the chain of
subsequent neutron-emitting nuclei, $B_f$ = 4.87 MeV, 4.98 MeV,
5.77 MeV in $^{302}$119, $^{301}$119 and $^{300}$119, while in a
chain of neutron decays starting from the $^{300}$119 nucleus, the
predicted fission barriers are 5.77 MeV, 5.55 MeV and 6.03 MeV,
respectively. Very recently Zagrebaev et al. \cite{Zagreb-12} have
reported a prediction for the same reaction,
$^{254}$Es($^{48}$Ca,xn)$^{302-x}$119 (about 0.3 pb for $3n$
channel). No prediction for the
$^{252}$Es($^{48}$Ca,xn)$^{300-x}$119 reaction was given.

In case of inaccessibility of Es targets, the most promising
target-projectile combination to synthesize the element $Z=119$ is
the $^{249}$Bk($^{50}$Ti,xn)$^{299-x}$119 reaction. Predictions
for this reaction are shown in Fig. 5(c). Both $3n$ and $4n$
channels are expected to have comparable cross sections of about
30 fb (at maximum) at $E_{c.m.} \approx 225$ and 232 MeV,
respectively. Almost an equally small cross section for the
$^{249}$Bk($^{50}$Ti,xn)$^{299-x}$119 reaction (about 60 fb) was
predicted in Ref. \cite{Zagreb-08}, and somewhat larger value
(about 110 fb) in Ref. \cite{Wang-12}. Unfortunately, such small
cross sections seem to be beyond the reach of present-state
experiments. More optimistic predictions for the same reaction
appeared recently in Ref. \cite{Liu-119}, however a relatively
large cross section (about 0.6 pb) was obtained for probably
overestimated values of the fission barrier taken as the pure
ground-state shell effect from tables of Ref. \cite{Moller}.

Prospects for the synthesis of element $Z$ = 120 are considerably
worse than those for $Z$ = 119. First of all, there is no chance
to use the favorable beam of $^{48}$Ca because the complementary
$^{257}$Fm target cannot be produced. We consider therefore
reactions with $^{50}$Ti beam on two available isotopes of
californium, $^{249}$Cf($^{50}$Ti,xn)$^{299-x}$120 and
$^{251}$Cf($^{50}$Ti,xn)$^{301-x}$120, which seem to be best
choice. Excitation functions for these two reactions are shown in
Figs. 5(d) and 5(e). The largest cross section is expected in the
former reaction (about 6 fb at maximum in both $3n$ and $4n$
channels), in the latter reaction the maximum cross section is
about 3 fb for $4n$ channel. Again, similarly as in case of
reactions on two isotopes of einsteinium discussed above, a
smaller cross section for more neutron rich compound nucleus is
associated with respectively lower fission barriers predicted in
Refs. \cite{Kowal-10,Kow-Sob}.

In Fig. 5(f) we present results of calculations for the
$^{248}$Cm($^{54}$Cr,xn)$^{302-x}$120 reaction that is a more
symmetric combination of even-$Z$ target and projectile, next to
Ti + Cf. The obtained cross sections of the order of 1 fb for $3n$
and $4n$ reaction channels clearly demonstrate that fusion
processes are too strongly hindered in more symmetric systems. For
completeness, we calculated also cross sections in two reactions
of much more symmetric systems,
$^{238}$U($^{64}$Ni,xn)$^{302-x}$120 and
$^{244}$Pu($^{58}$Fe,xn)$^{302-x}$120 (not shown in figures), for
which attempts to produce the element $Z$ = 120 were done
\cite{GSI-120}, \cite{Dubna-120}. The calculated $3n$ and $4n$
cross sections in these two reactions are dramatically small,
about 0.3 fb and 0.1 fb, respectively. Note that experimental
upper limits for these two reactions had been established at 90 fb
\cite{GSI-120} and 400 fb \cite{Dubna-120}, respectively.

Our calculations show that if the fission barriers of Refs.
\cite{Kowal-10,Kow-Sob} were correct, there is no chance to
synthesize the element $Z$ = 120, even in the most favorable
reaction $^{249}$Cf($^{50}$Ti,xn)$^{299-x}$120, for which the
predicted cross section is only 6 fb. Note that other model
calculations for the $^{249}$Cf($^{50}$Ti,xn)$^{299-x}$120
reaction, published previously \cite{Zagreb-08,Liu-120,Siwek-120,
Nasirov-120,Nasirov-120a,Wang-12}, predicted considerably larger
cross sections though also too small to be measurable (typically
of the order of 50 fb). The dispersion of these different
theoretical results has to be linked, first of all, to different
fission barriers and ground-state masses used in these
calculations.

We would like to emphasize that our predictions concerning the
synthesis of $Z=119$ and 120 nuclei are based on the {\em
consistency} of the FBD model calculations with the adopted
ground-state {\em masses} and {\em fission barriers} of Refs.
\cite{Kowal-10,Kow-Sob} and with {\em all} the existing {\em
experimental data} on the synthesis of superheavy nuclei in hot
fusion reactions
\cite{Oganessian,Oga-243Am,Oga-244Pu,Oga-242Pu,Oga-249Cf,LBL-242Pu1,Oga-249Bk,GSI-244Pu,LBL-242Pu2,GSI-248Cm,GSI-244Pu2,Oga-243Am2}.
Therefore the accuracy of these predictions is expected to be
comparable with the accuracy of our overall fit to the data for
the synthesis of $Z$ = 114--118 nuclei, shown in Fig. 2.

In summary, we analyzed a complete set of existing data on hot
fusion reactions leading to the synthesis of superheavy nuclei of
$Z$ =114-118
\cite{Oganessian,Oga-243Am,Oga-244Pu,Oga-242Pu,Oga-249Cf,LBL-242Pu1,Oga-249Bk,GSI-244Pu,LBL-242Pu2,GSI-248Cm,GSI-244Pu2,Oga-243Am2}
 in terms of a new $l$-dependent version of the FBD
model  with fission barriers and ground-state masses taken from
the macroscopic-microscopic model of Kowal et al.
\cite{Kowal-10,Kow-Sob}. By ``calibrating'' the assumed
injection-point distances ($s_{inj}$) to the measured cross
sections, perfect systematics of $s_{inj}$-values have been
established for a wide range of hot fusion reactions enabling,
hopefully, reliable predictions of the synthesis cross sections
for yet unexplored reactions. Regarding prospects to produce the
new element $Z$ = 119, our calculations prefer the
$^{252}$Es($^{48}$Ca,xn)$^{300-x}$119 reaction, for which the
synthesis cross section of about 0.2 pb in $4n$ channel at
$E_{c.m.}\approx 220$ MeV is expected. According to the
microscopic-macroscopic model predictions \cite{Kowal-10,Kow-Sob},
fission barriers for heavier isotopes of the element $Z$ = 119 are
significantly lower leading to a considerably smaller cross
section in the alternative $^{254}$Es($^{48}$Ca,xn)$^{302-x}$119
reaction. Also the reaction $^{249}$Bk($^{50}$Ti,xn)$^{299-x}$119
gives little chances for a measurable cross section (the predicted
cross section is about 30 fb for both $3n$ and $4n$ channels). The
most favorable reaction to synthesize the element $Z$ = 120 is the
$^{249}$Cf($^{50}$Ti,xn)$^{299-x}$120 reaction, but the predicted
cross section is only 6 fb (for $3n$ and $4n$ channels).

\begin{figure}[p]
\includegraphics[width=10cm,angle=0]{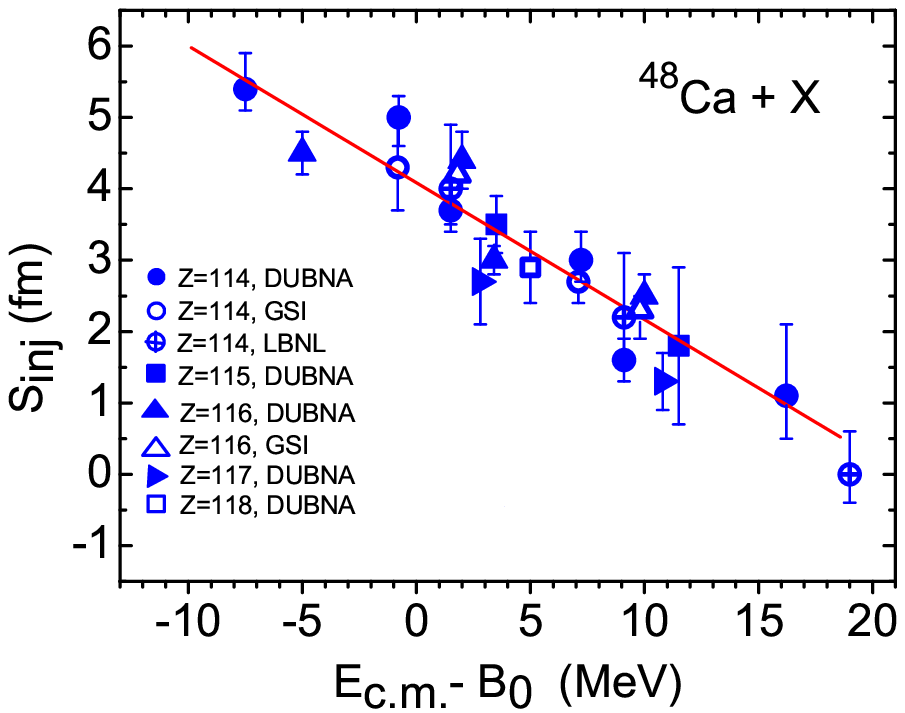}
\caption{(Color online) Systematics of the injection-point
distance $s_{inj}$ as a function of the kinetic energy excess
$E_{c.m.}-B_0$ above the Coulomb barrier $B_0$. Values of
$s_{inj}$ have been determined for each reaction and each
particular $xn$ channel by fitting the theoretical cross section
at the maximum of a given $xn$ excitation function to the data.
The calculations have been done for the fission barrier heights
and ground-state masses of Kowal et al. \cite{Kowal-10,Kow-Sob}.
Complete list of the analyzed reactions with references is given
in the text. Identical symbols for a given $Z$ and a given
experiment refer to data for consecutive $xn$ channels.}
\end{figure}

\begin{figure}[p]
\begin{tabular}{cc}
\resizebox{70mm}{!}{\includegraphics[angle=0,scale=0.75]{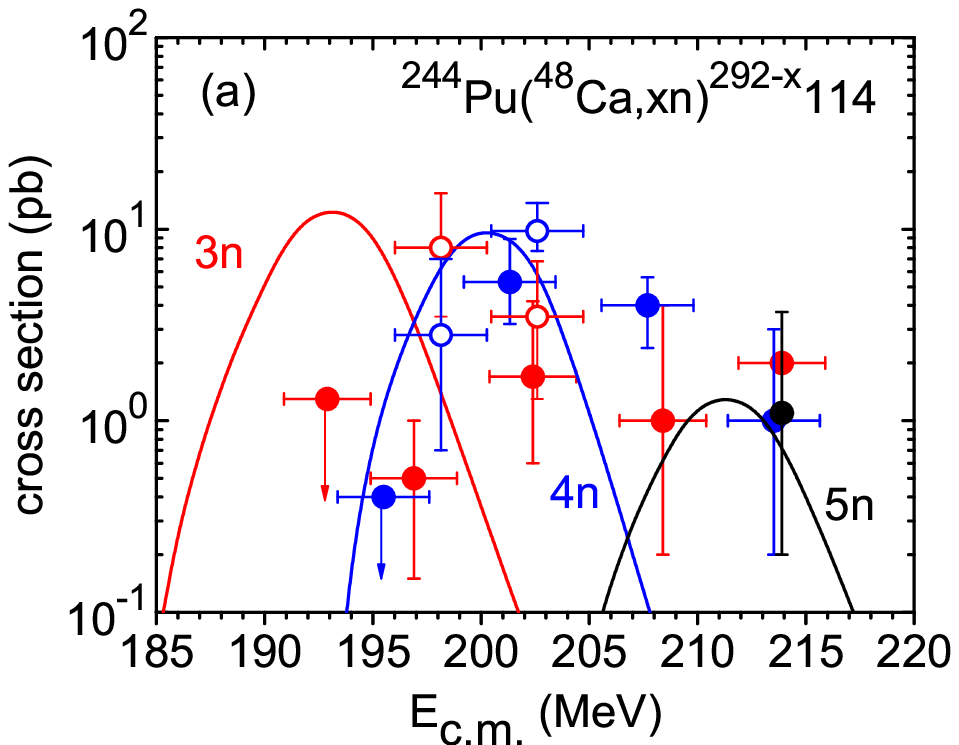}}
&
\resizebox{70mm}{!}{\includegraphics[angle=0,scale=0.75]{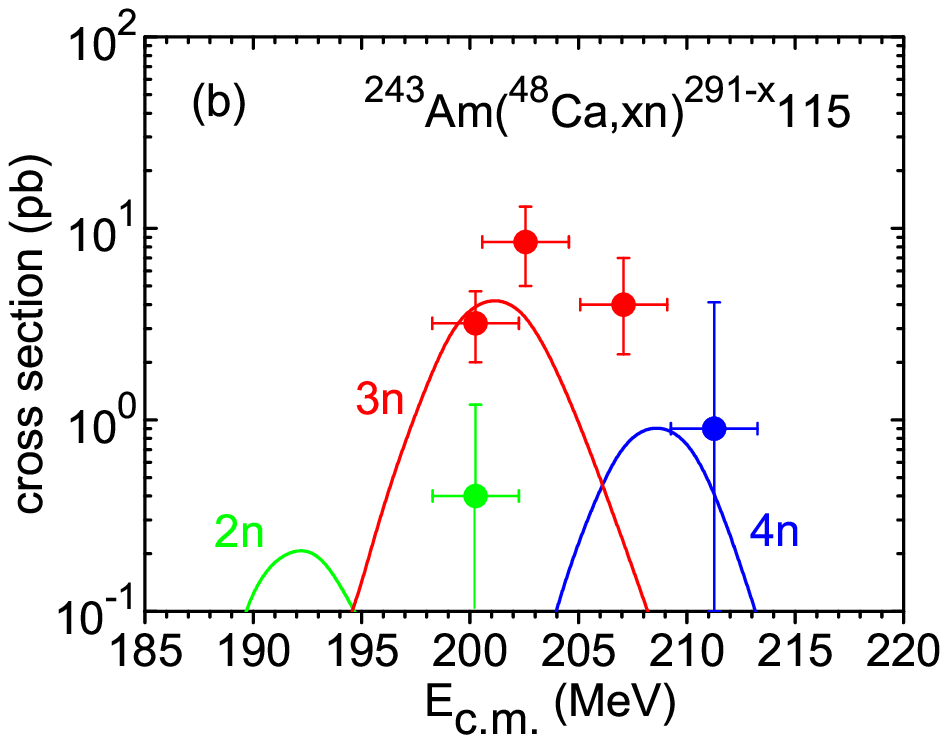}}\\

\resizebox{70mm}{!}{\includegraphics[angle=0,scale=0.75]{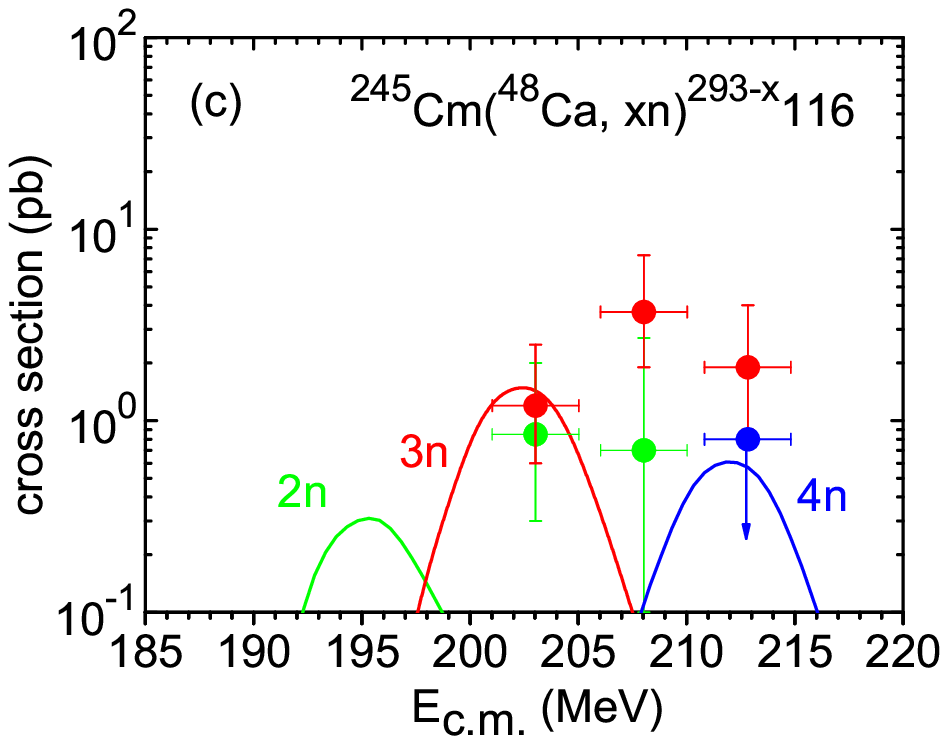}}
&
\resizebox{70mm}{!}{\includegraphics[angle=0,scale=0.75]{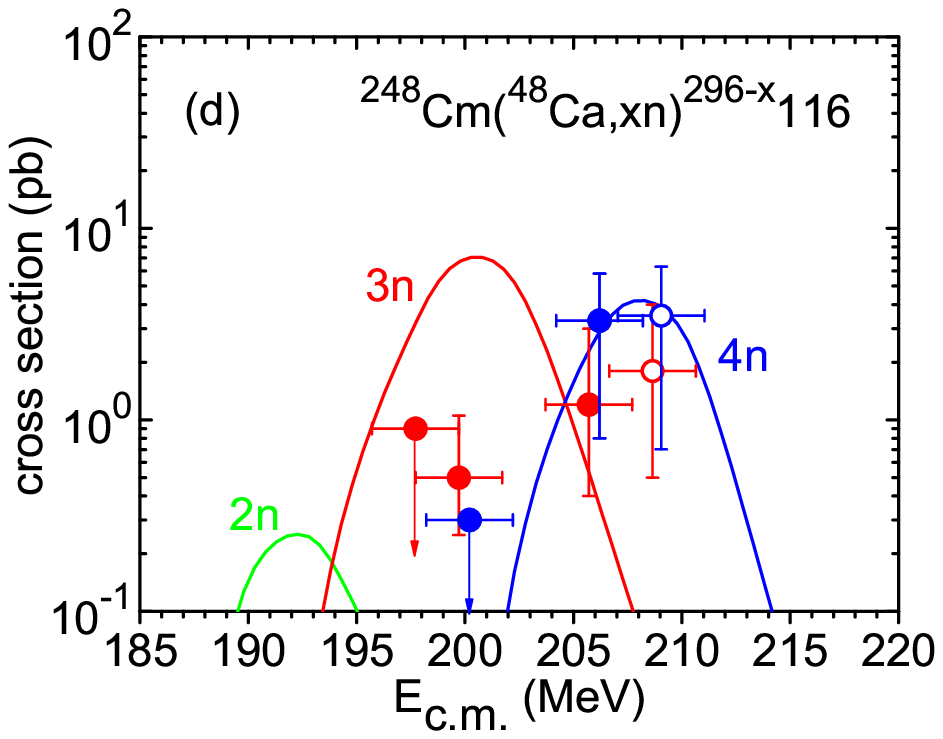}}\\

\resizebox{70mm}{!}{\includegraphics[angle=0,scale=0.75]{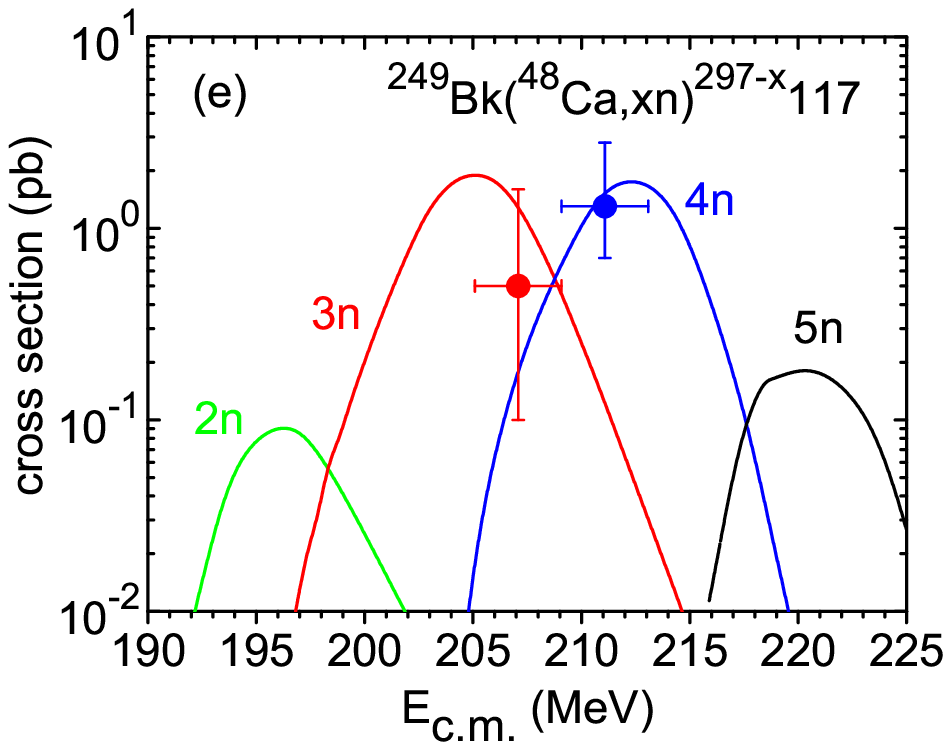}}
&
\resizebox{70mm}{!}{\includegraphics[angle=0,scale=0.75]{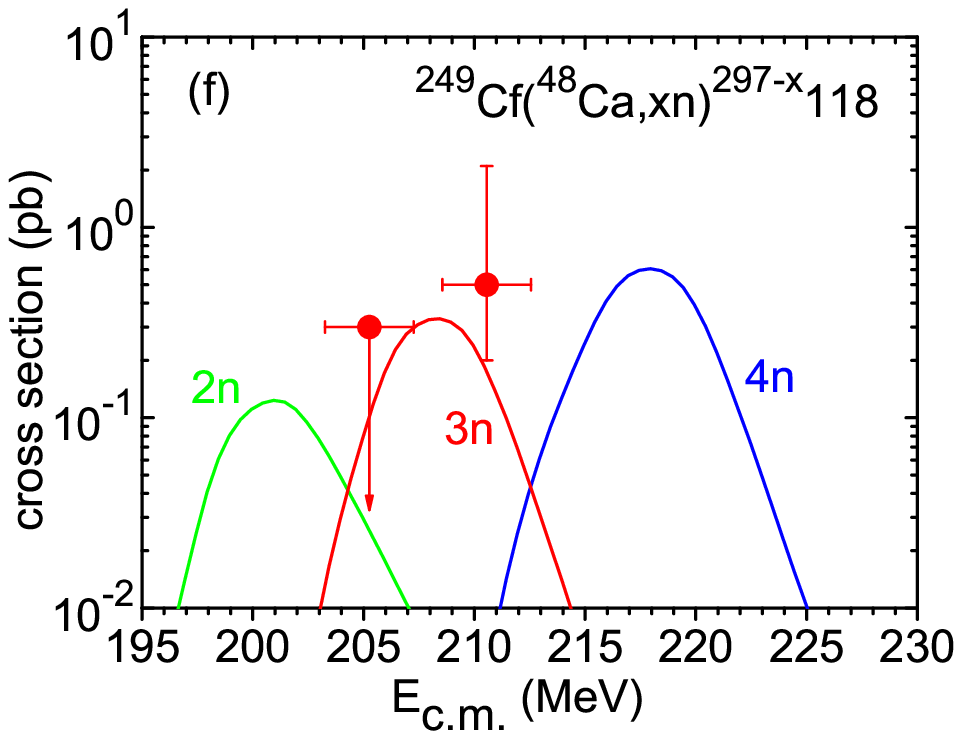}}\\

\end{tabular}
\caption{(Color online) Energy dependence of the cross section for
synthesis of superheavy nuclei in hot fusion reactions. Full
circles represent data for $3n$, $4n$ and $5n$ reaction channels
obtained in Dubna experiments for elements $Z$ = 114--118
\cite{Oganessian,Oga-244Pu,Oga-242Pu,Oga-243Am,Oga-243Am2,Oga-249Bk,Oga-249Cf};
open circles represent data obtained at GSI Darmstadt for $Z$ =
114 and 116 \cite{GSI-244Pu,GSI-248Cm}. Data are compared with
excitation functions for separate $xn$ channels, calculated with
the FBD model assuming fission barriers and ground-state masses of
Kowal et al. \cite{Kowal-10,Kow-Sob} and the systematics of the
injection-point distance, Eq. (\ref{sinj}).}

\end{figure}

\begin{figure}[p]
\includegraphics[width=10cm,angle=0]{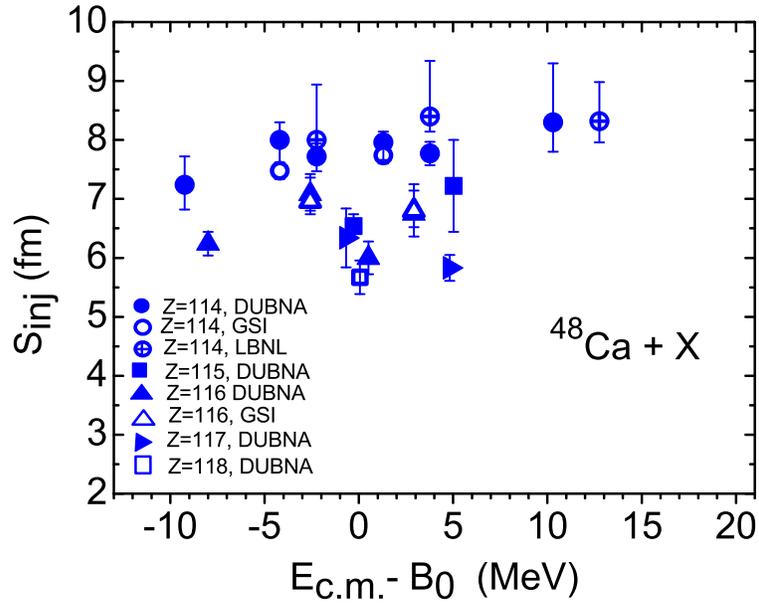}
\caption{(Color online) Dependence of the injection-point distance
$s_{inj}$ on the kinetic energy excess $E_{c.m.}-B_0$ above the
Coulomb barrier $B_0$, deduced from analysis of experimental data
\cite{Oganessian,Oga-243Am,Oga-244Pu,Oga-242Pu,Oga-249Cf,LBL-242Pu1,Oga-249Bk,GSI-244Pu,LBL-242Pu2,GSI-248Cm,GSI-244Pu2,Oga-243Am2}
the same way as in Fig. 1, but assuming the fission barrier
heights \cite{Moller-barriers} and ground-state masses
\cite{Moller} of M\"oller et al. See text. }
\end{figure}

\begin{figure}[p]
\includegraphics[width=10cm,angle=0]{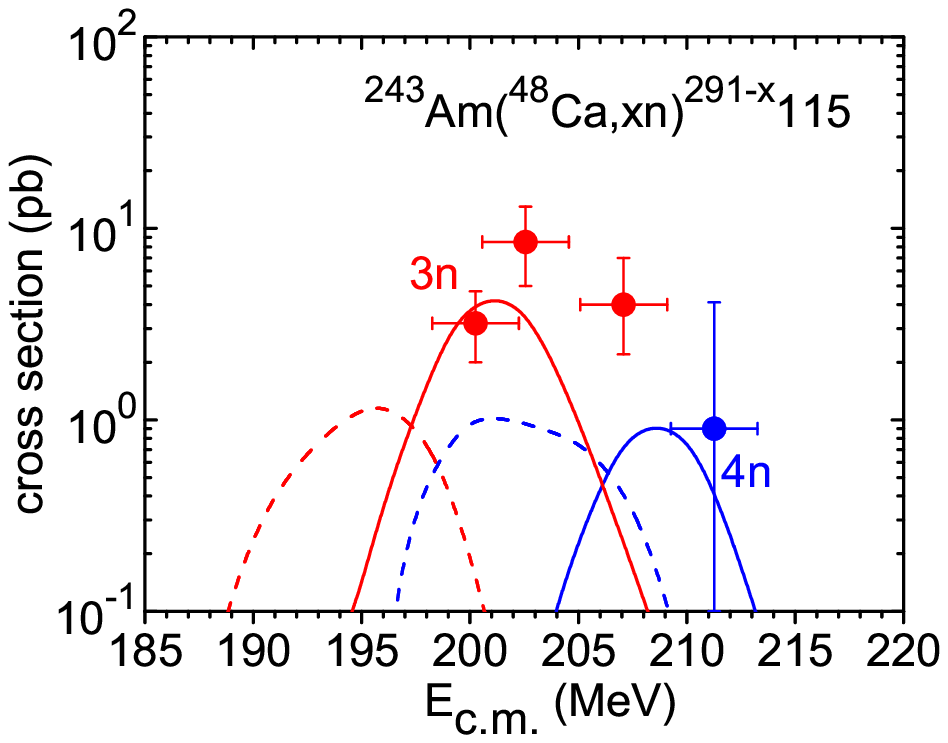}
\caption{(Color online) Excitation functions for the 3n and 4n
channels of the $^{243}$Am($^{48}$Ca,xn)$^{291-x}$115 reaction
calculated with the FBD model assuming the fission barriers
\cite{Moller-barriers} and ground-state masses \cite{Moller} of
M\"oller et al. (dashed lines) compared with the experimental
cross sections \cite{Oganessian,Oga-243Am,Oga-243Am2} and the
predictions for the fission barriers and ground-state masses of
Kowal et al. \cite{Kowal-10,Kow-Sob} (solid lines). In the absence
of clear correlation between $s_{inj}$ and $E_{c.m.}-B_0$ for the
barriers of M\"oller et al. (see Fig. 3), the dashed lines were
calculated for a fixed value $s_{inj}$ = 7.2 fm (the mean value).
}
\end{figure}

\begin{figure}[p]
\begin{tabular}{cc}
\resizebox{70mm}{!}{\includegraphics[angle=0,scale=0.75]{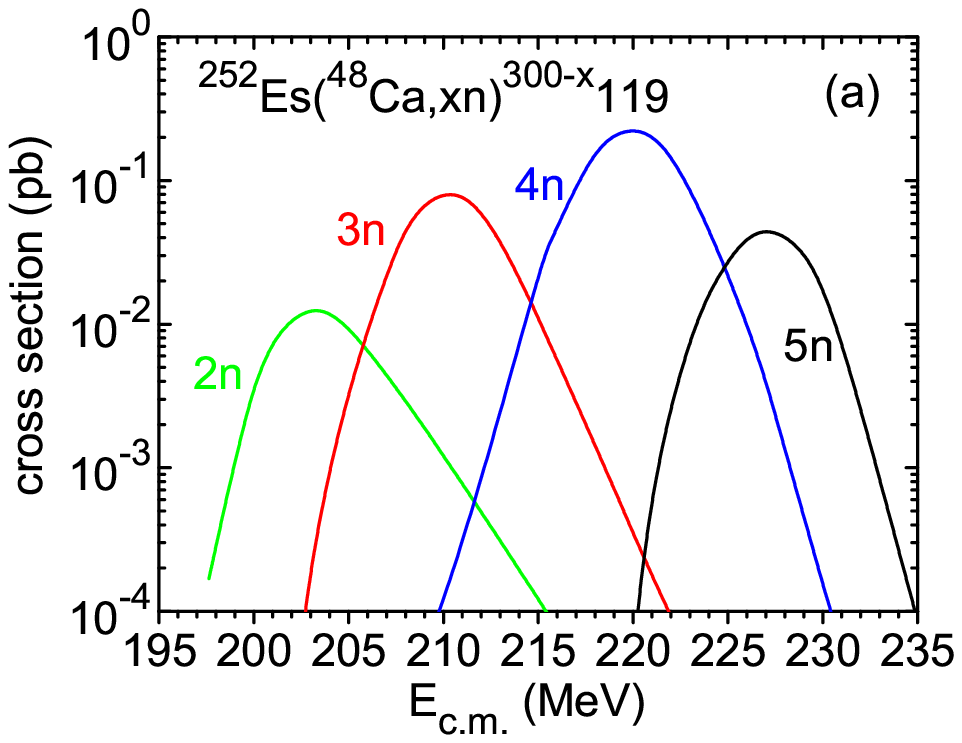}}
&
\resizebox{70mm}{!}{\includegraphics[angle=0,scale=0.75]{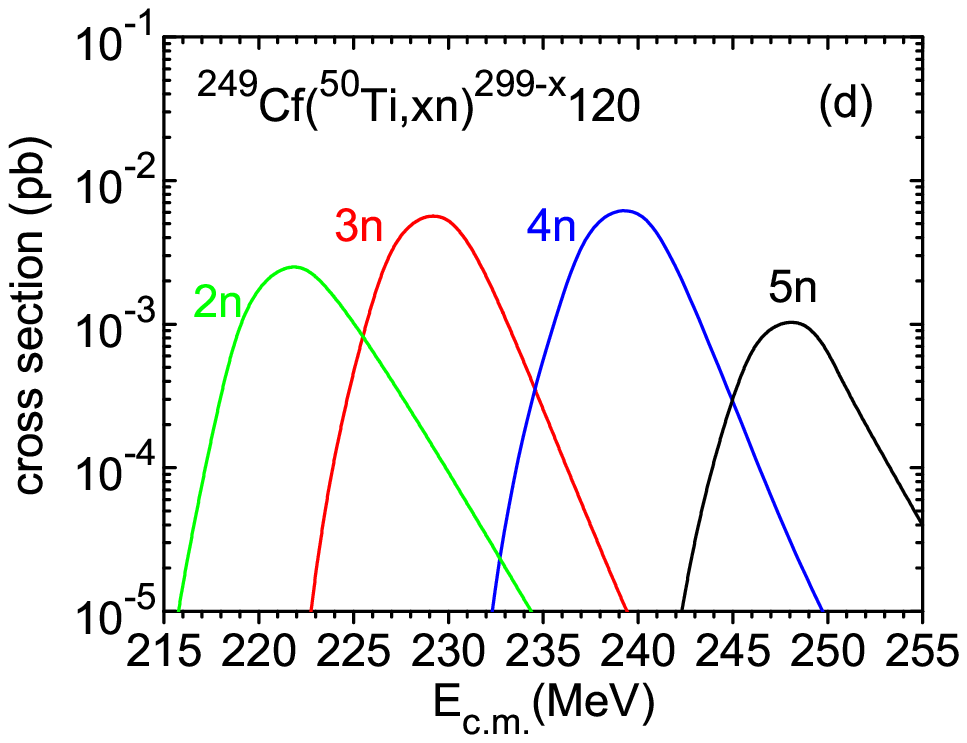}}\\

\resizebox{70mm}{!}{\includegraphics[angle=0,scale=0.75]{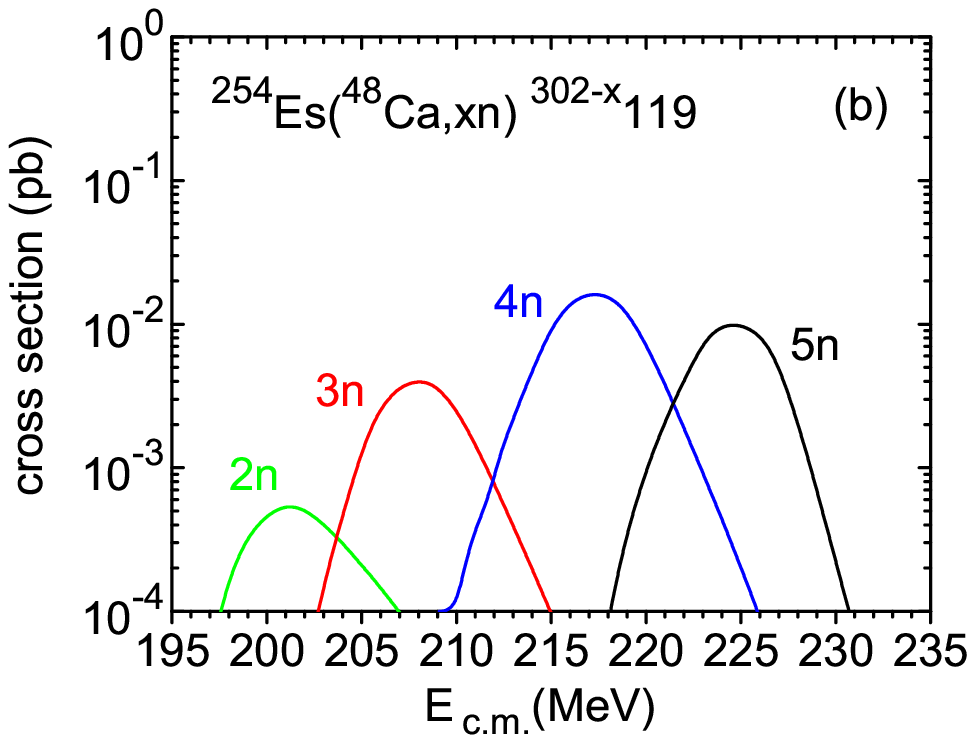}}
&
\resizebox{70mm}{!}{\includegraphics[angle=0,scale=0.75]{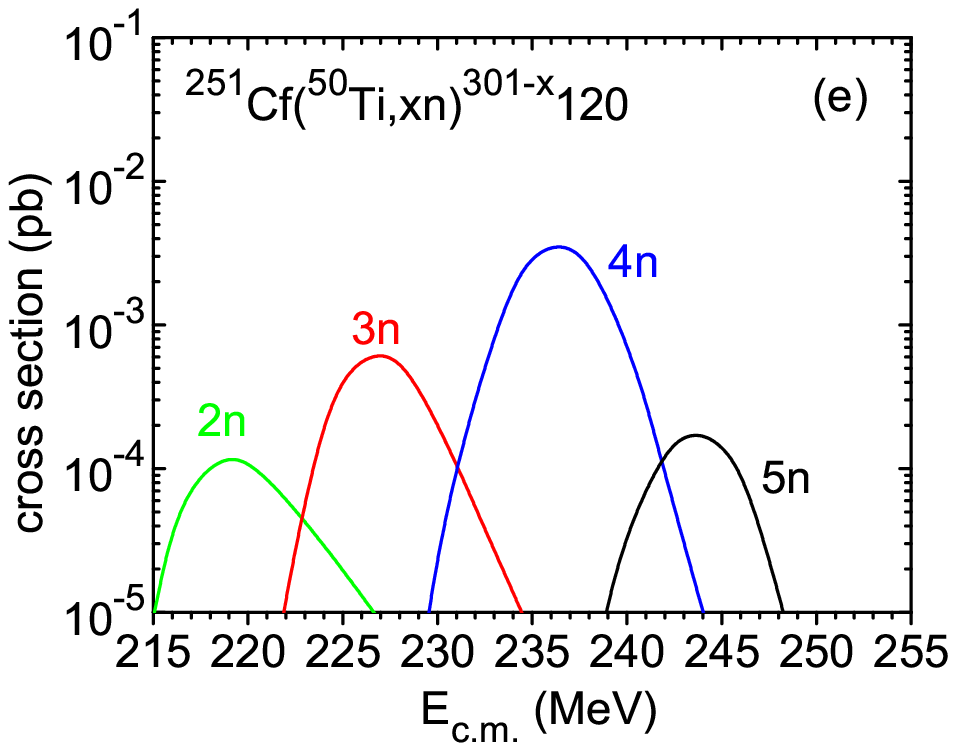}}\\

\resizebox{70mm}{!}{\includegraphics[angle=0,scale=0.75]{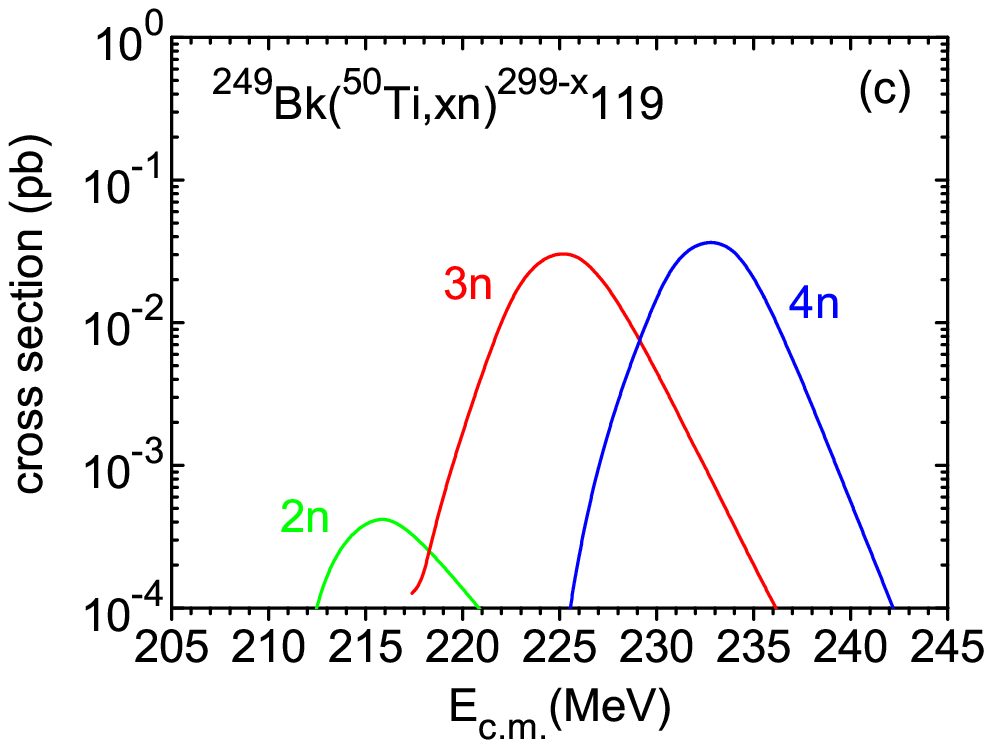}}
&
\resizebox{70mm}{!}{\includegraphics[angle=0,scale=0.75]{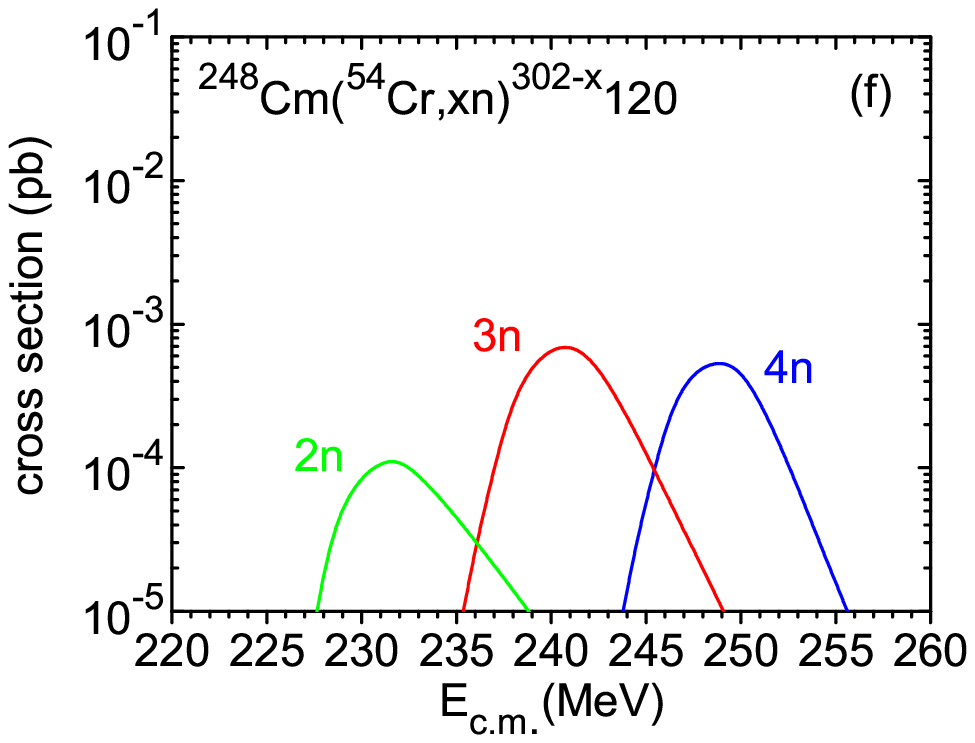}}\\

\end{tabular}
\caption{(Color online) Synthesis cross sections of undiscovered
yet superheavy nuclei of $Z$ = 119 and 120 predicted by using the
fusion-by-diffusion (FBD) model with fission barriers and
ground-state masses of Kowal et al. \cite{Kowal-10,Kow-Sob} and
the systematics of the injection-point distance, Eq. (\ref{sinj}).
See text.}
\end{figure}

\end{document}